\documentclass{PoS}
\usepackage{amssymb}

\title{Finite-size effects and the search for the critical endpoint of QCD}

\ShortTitle{Finite-size effects and the search for the critical endpoint of QCD}

\author{\speaker{Eduardo Souza Fraga}\\%
       Instituto de F\'\i sica, Universidade Federal do Rio de Janeiro, \\
       Caixa Postal 68528, Rio de Janeiro, RJ 21941-972, Brazil\\
       E-mail: \email{fraga@if.ufrj.br}}
       
\author{Takeshi Kodama\\
        Instituto de F\'\i sica, Universidade Federal do Rio de Janeiro, \\
       Caixa Postal 68528, Rio de Janeiro, RJ 21941-972, Brazil\\ 
       E-mail: \email{tkodama@if.ufrj.br}}
       
\author{Let\'\i cia F. Palhares\\
        Instituto de F\'\i sica, Universidade Federal do Rio de Janeiro, \\
       Caixa Postal 68528, Rio de Janeiro, RJ 21941-972, Brazil\\ 
       Institut de Physique Th\'eorique, CEA/DSM/Saclay, \\ Orme des Merisiers, 
       91191 Gif-sur-Yvette cedex, France \\
       E-mail: \email{leticia@if.ufrj.br}}
       
\author{Paul Sorensen\\
       Physics Department, Brookhaven National Laboratory, \\
       Upton, NY 11973-5000, USA\\
       E-mail: \email{psoren@bnl.gov}}

\abstract{Taking into account the finiteness of the system created in heavy ion collisions, we show sizable results for the modifications of the chiral phase diagram at volume scales typically encountered in current experiments and demonstrate the applicability of finite-size scaling as a tool in the experimental search for the critical endpoint. Using data from RHIC and SPS and assuming finite-size scaling, we find that RHIC data from $200~$GeV down to $19.6~$GeV is only consistent with a critical point at $\mu\gtrsim 510~$MeV. We also present predictions for the fluctuations at lower energies currently being investigated in the Beam Energy Scan program.}

\FullConference{The many faces of QCD\\
		 November 1-5, 2010\\
		 Gent, Belgium}

\begin{document}

Relativistic heavy ion collision experiments have entered a new era of investigations of the phase 
diagram of strong interactions. The Beam Energy Scan program at RHIC \cite{BES} is probing a 
relatively wide window in center-of-mass energy, which translates into higher values of the baryonic 
chemical potentials and lower values of the temperature associated with the plasma created in 
the collision process. The hope is that the scan will able to reach the region where a second-order 
critical point is expected to exist \cite{critical}. Therefore, pragmatic theoretical tools that can be 
phenomenologically applied to data analysis are demanded.

Identifying the presence of a critical point, and even of a first-order transition coexistence 
line, is not a simple task in heavy ion experiments. One has to uncover, from different sorts of complex 
backgrounds, signatures related to large fluctuations that would stem from the critical behavior of 
the order parameter of the chiral transition, the chiral condensate. In a system in equilibrium and in 
the thermodynamic limit, those are a consequence of the unlimited growth of the correlation 
length \cite{Cardy:1996xt}, and in the case of heavy ions would lead to non-monotonic 
behavior \cite{Stephanov:1998dy,Stephanov:2008qz} or sign modifications \cite{Asakawa:2009aj} 
of particle correlation fluctuations. 

Although often compared to the case of the early universe at the time of the primordial quark-hadron 
transition, the space-time scales present in the dynamics of the quark-gluon plasma formed in heavy 
ion collisions differ from the cosmological ones by almost twenty orders of magnitude. So, realistically, 
this system is usually small, short-lived, and part of the time out of equilibrium. The finite (short) lifetime 
of the plasma state and critical slowing down could severely constrain the growth of the correlation 
length, as shown by estimates in Refs. \cite{Berdnikov:1999ph,Stephanov:2009ra}. The finite size of 
the system could dramatically modify the phase structure of strong interaction, as shown using lattice 
simulations \cite{Gopie:1998qn,Bazavov:2007zz} and different effective model 
approaches \cite{finite-NJL,Braun:2004yk,Yamamoto:2009ey,Palhares:2009tf}, 
and also affect significantly the dynamics of phase conversion \cite{Spieles:1997ab,Fraga:2003mu}.
As a consequence, all signatures of criticality based on non-monotonic behavior of particle 
correlation fluctuations will probe a \textit{pseudocritical point} that can be significantly shifted from 
the genuine (unique) critical endpoint by finite-size corrections and will be sensitive to boundary 
effects.

\begin{figure}[tbh]
\vspace{0.5cm}
\par
\begin{minipage}[t]{75mm}
\includegraphics[width=7cm]{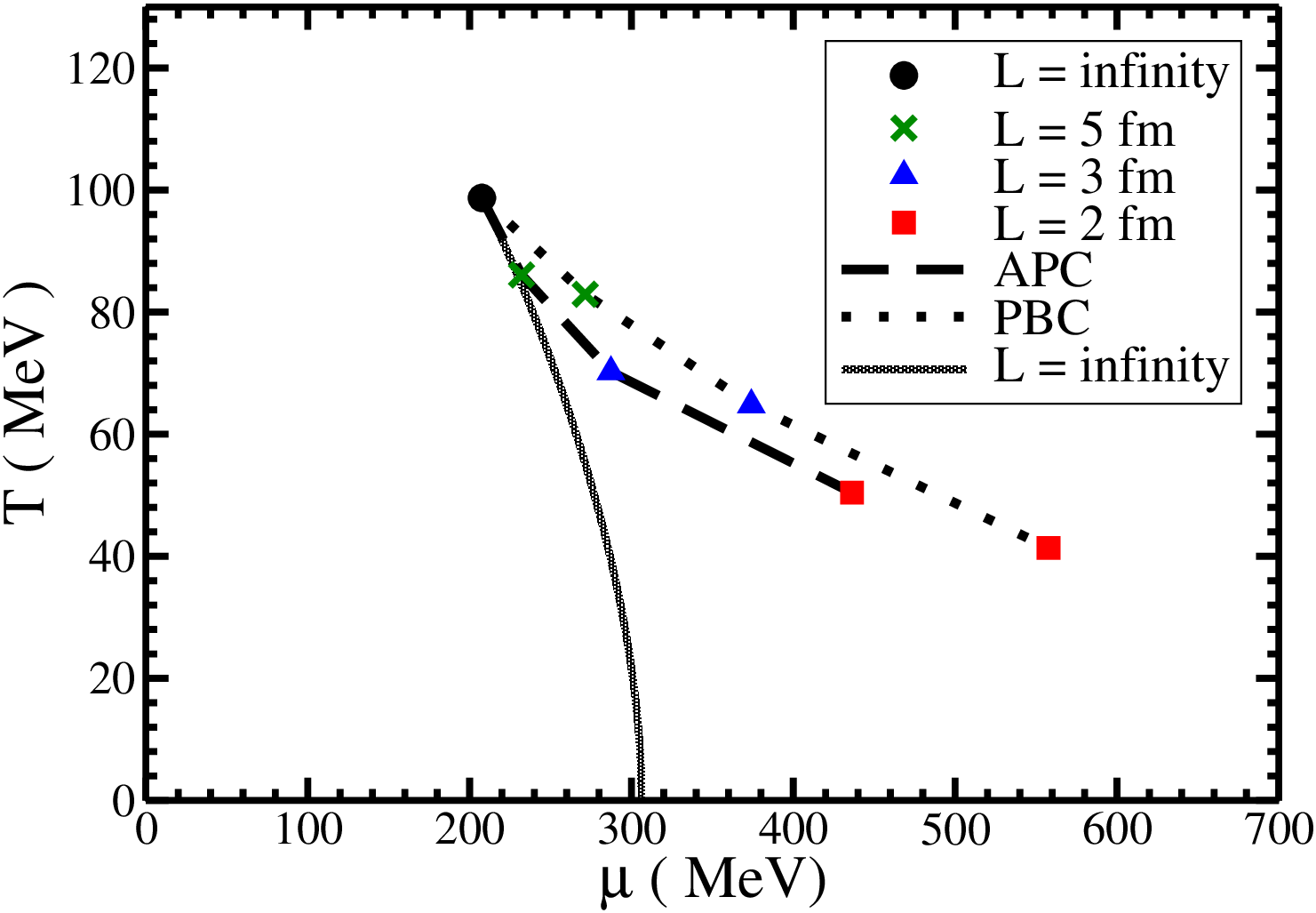}
\label{CEP}
\caption{Displacement of the pseudocritical endpoint in the $T-\mu$ plane as the system size is
decreased for different boundary conditions.}
\end{minipage}
\hspace{.3cm} 
\begin{minipage}[t]{75mm}
\includegraphics[width=7.3cm]{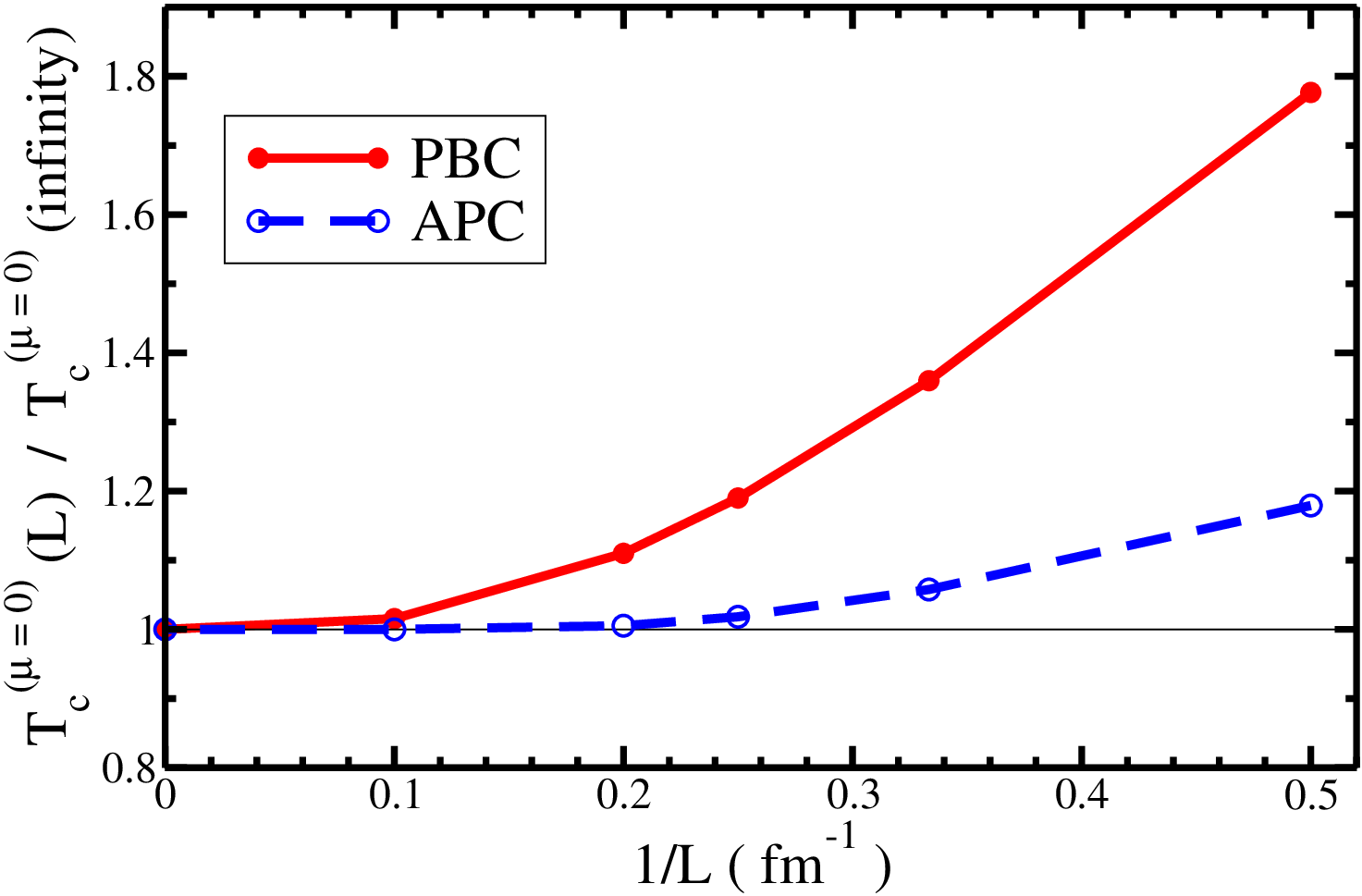}
\label{Tcrossover}
\caption{Normalized crossover temperature at $\mu=0$ as a function of the inverse size $1/L$ for the cases with PBC and APC.}
\end{minipage}
\end{figure}

The latter effect was demonstrated using the linear sigma model coupled to quarks with two 
flavors \cite{GellMann:1960np} as an effective theory for the chiral transition in 
Ref. \cite{Palhares:2009tf}. Here we illustrate the relevance the effects coming from the finite 
size of the system, of typical linear size $L$,  and the nature of the boundary might have in 
the investigation of the phase diagram of strong interactions using heavy ion collisions in Figs. 1 
and 2. 
Fig. 1 shows the displacement of the pseudocritical point, comparing periodic boundary 
conditions (PBC) and anti-periodic boundary conditions (APC): both coordinates of the critical 
point are significantly modified, and $\mu_{\mathrm{CEP}}$ is about $30\%$ larger for PBC. 
For $\mu=0$, the crossover transition is also affected by finite-size corrections, increasing as the 
system decreases, as shown in Fig. 2. Again, PBC generate larger effects: up to 
$\sim80\%$ increase in the crossover transition temperature at $\mu=0$ when $L=2$ fm. 
The range of values for the linear size $L$ is motivated by the estimated plasma
size presumably formed in high-energy heavy ion collisions at RHIC \cite{Adams:2005dq}.
The upper limit is essentially geometrical, provided by the radius of the nuclei involved,
whereas the lower limit is an estimate for the smallest plasma observed.

Nevertheless, the finiteness of the system in heavy ion collisions also brings a bright side: the possibility 
of using finite-size scaling (FSS) analysis \cite{fisher,Brezin:1981gm,Brezin:1985xx}. FSS is a 
powerful statistical mechanics technique that prescinds from the knowledge of the details of a 
given system; instead, it provides information about its criticality based solely on very general 
characteristics. And since the thermal environment corresponding to the region of quark-gluon 
plasma formed in heavy ion collisions can be classified according to the centrality of the collision, 
events can be separated according to the size of the plasma that is created. So, heavy ion collisions indeed provide an ensemble of differently-sized systems.

Although it is clearly not simple to define an appropriate scaling variable in the case of heavy ion 
collisions, the flexibility of the FSS method allows for a pragmatic approach for the use of scaling plots 
in the search for the critical endpoint as was delineated in Ref. \cite{Palhares:2009tf} and performed 
in Ref. \cite{Fraga:2011hi}. The essential point is that although the reduced volume of the plasma 
formed in high-energy heavy ion collisions will dissolve a possible critical point into a region and 
make the effects from criticality severely smoothened, as discussed above, the non-monotonic 
behavior of correlation functions for systems of different sizes, given by different centralities, must 
obey FSS near criticality \cite{Cardy:1996xt,amit}. 

\begin{figure}[tbh]
\vspace{0.5cm}
\par
\begin{minipage}[t]{75mm}
\includegraphics[width=7.3cm]{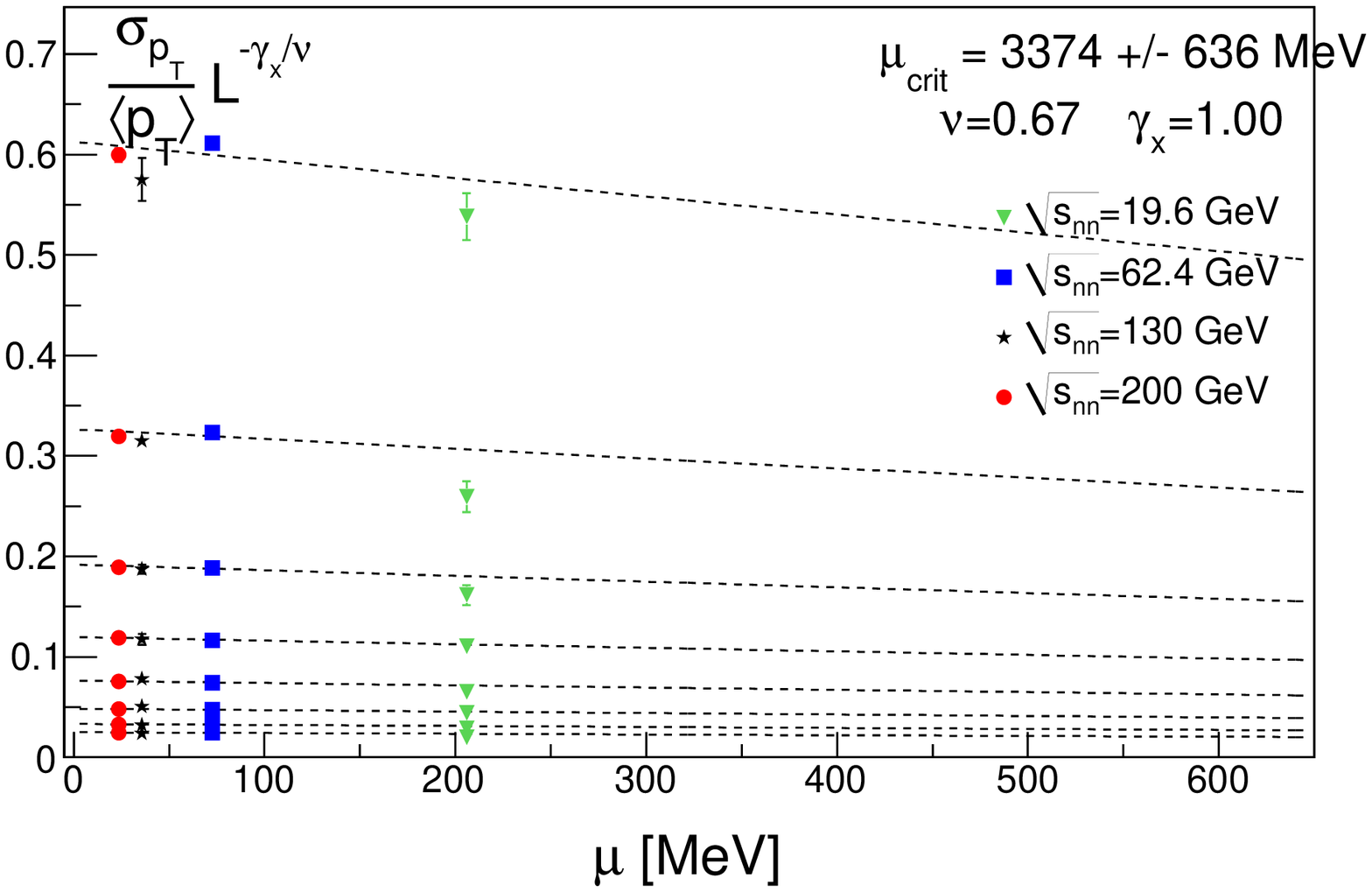}
\caption{Scaled $\sigma_{p_T}/\langle p_T\rangle$ vs $\mu$ for different system sizes, and 
with $\nu=2/3$ and $\gamma_{x}=1$. Data extracted from RHIC collisions at energies 
$\sqrt{s_{NN}}=19.6, 62.4, 130$, and $200$ GeV (linear fit, see text).}
\label{scaling-linear}
\end{minipage}
\hspace{.3cm} 
\begin{minipage}[t]{75mm}
\includegraphics[width=7.3cm]{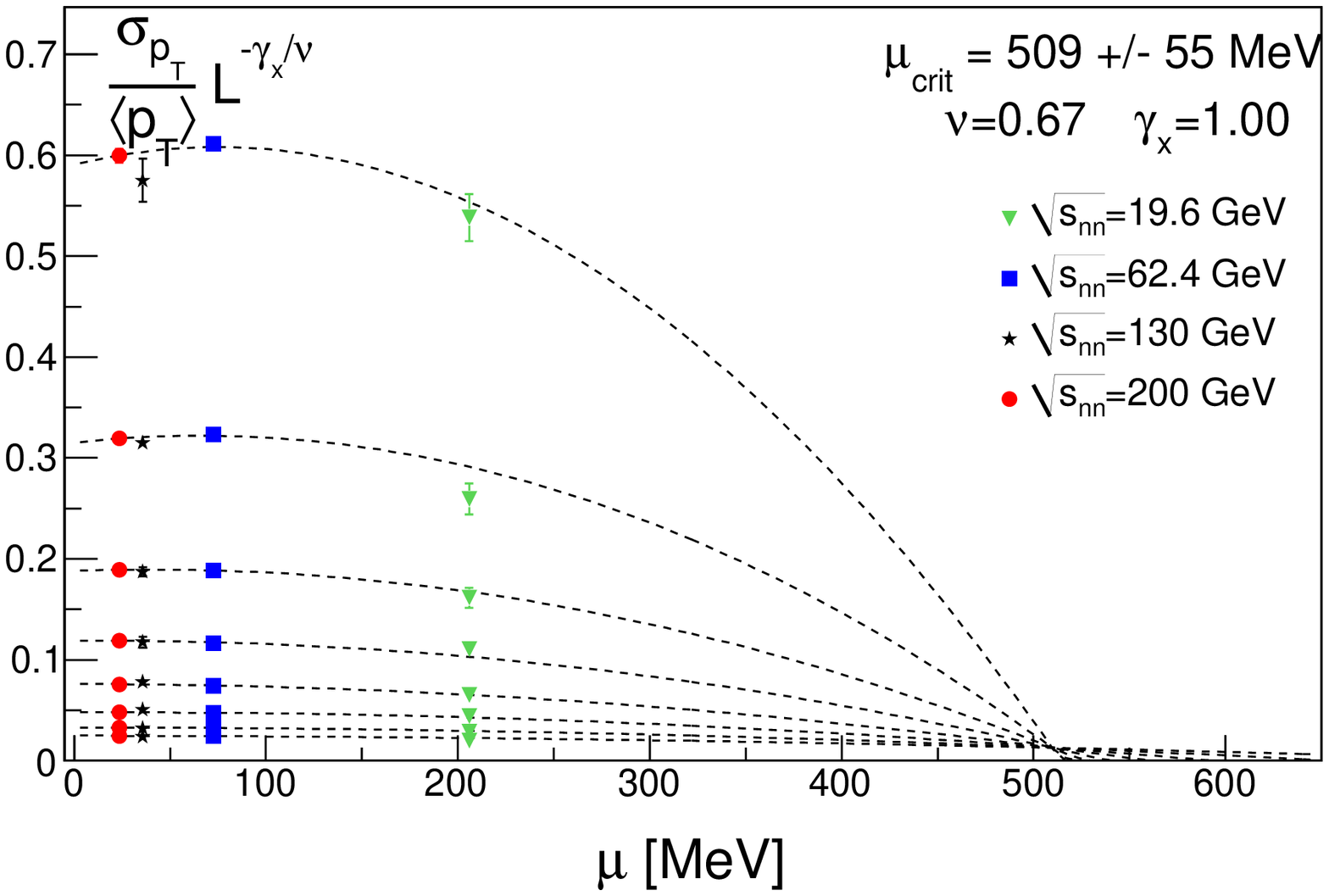}
\caption{Scaled $\sigma_{p_T}/\langle p_T\rangle$ vs $\mu$ 
for different system sizes. Again, $\nu=2/3$ and $\gamma_{x}=1$.
Data extracted from RHIC collisions at energies $\sqrt{s_{NN}}=19.6, 62.4, 130$, and $200$ GeV 
(second order polynomial fit, see text).}
\label{scaling-pol2}
\end{minipage}
\end{figure}

Using data from RHIC and SPS and defining appropriate scaling variables, we can generate scaling 
plots for data sets with $\sqrt{s_{NN}}=17.3, 19.6, 62.4, 130, 200$ GeV. For finite $L$, crossover 
effects become important. If the correlation length diverges as $\xi_{\infty}\sim t^{-\nu}$ at criticality, 
where $\nu$ is the corresponding critical exponent, in the case of $L^{-1} t^{-\nu} \gg 1$ the system 
is no longer governed by the critical fixed point and  $L$ limits the growth of the correlation length, 
rounding all singularities \cite{Cardy:1996xt,amit}. If $L$ is finite, $\xi$ is analytic in the limit $t \to 0$, 
and one can draw scaling plots of $L/\xi$ vs. some coupling for different values of $L$ to find that all curves cross at a given value in this limit, which is a way to determine its critical value. The critical 
temperature and so on can also be determined in this fashion, since the curves will also cross 
at $t=t_{c}$. This scaling plot technique can be extended, taken to its full power for other quantities, 
such as correlation functions. An observable $X$ in a finite thermal system can be written, 
in the neighborhood of criticality, in the following form \cite{Brezin:1985xx}:
\begin{equation}
X(t,L)=L^{\gamma_{x}/\nu} f(t L^{1/\nu}) \;,  \label{scaling}
\end{equation}
where $\gamma_{x}$ is the bulk (dimension) exponent of $X$ and $\{g\}$ dimensionless coupling 
constants. The function $f(y)$ is universal up to scale fixing, and the critical exponents are sensitive 
essentially to dimensionality and internal symmetry, which will give rise to the different universality 
classes \cite{amit}. Using the appropriate scaling variable, all curves should collapse into one single 
curve if one is not far from the critical point. So, this technique can be applied to the analysis of 
observables  that are directly related to the correlation function of the order parameter of the transition, 
such as fluctuations of the multiplicity of soft pions \cite{Stephanov:1998dy}. The correct scaling 
variable should measure the distance from the critical point, thereby involving both temperature and 
chemical potential in the case of the QCD phase diagram. This would produce a two-dimensional 
scaling function and make the analysis of heavy ion data highly nontrivial. Phenomenologically, we 
adopt a simplification motivated by results from thermal models for the freeze-out region, connecting 
temperature and chemical potential. We can parametrize the freeze-out curve by $\sqrt{s_{NN}}$, 
and build our one-dimensional scaling variable from this quantity and the size of the system. For 
details, we refer the reader to Ref. \cite{Fraga:2011hi}.

To search for scaling, we consider the correlation measure 
$\sigma_{p_T}/\langle p_T\rangle$~\cite{Adams:2005ka} 
scaled by $L^{-\gamma_x/\nu}$, according to Eq. \ref{scaling}. We consider the $p_T$ fluctuations 
$\sigma_{p_T}$ scaled by $\langle p_T\rangle$ to obtain a dimensionless variable. We use the 
correlation data measured in bins corresponding to the $0-5$\%, $5-10$\%, $10-20$\%, 
$20-30$\%, $30-40$\%, $40-50$\%, $50-60$\%, and $60-70$\% most central collisions. 
We estimate the corresponding lengths $L$ to be $12.4$, $11.1$, $9.6$, $8.0$, $6.8$, $5.6$, 
$4.5$, and $3.4$ fm. The exponent $\nu = 2/3$ is determined by the Ising universality 
class of QCD and we consider values of $\gamma_{x}$ around $1$ (ignoring small anomalous 
dimension corrections). We also varied the value of $\gamma_{x}$ from $0.5$ to $2.0$ and found 
that changing $\gamma_{x}$ within this range does not improve the scaling behavior.

In Figs.~3 and 4 we plot $\sigma_{p_T}/\langle p_T\rangle$ scaled by 
$L^{-\gamma_x/\nu}$ vs $\mu$ for different system sizes, using data extracted from collisions at 
$\sqrt{s_{NN}}= 19.6, 62.4, 130, 200$ GeV. If there is a critical point at $\mu=\mu_{\mathrm{crit}}$, the 
curves for different sizes of the system should cross at this value of $\mu$. However, since the 
currently available data is restricted to not so large values of the chemical potential, one has to 
perform extrapolations using fits. The scaling function $f$ in Eq. (\ref{scaling}) is expected to be 
smoothly varying around the critical point, so we fit the data corresponding to a given linear size $L$ to 
a polynomial, but constraining the polynomials to enforce the condition that all the curves cross at 
some $\mu=\mu_{\mathrm{crit}}$, where $\mu_{\mathrm{crit}}$ is an adjustable parameter in the fit. 
This clearly assumes 
the existence of a critical point. In Fig.~3 we use a linear fit. The approximate energy independence of $\sigma_{p_T}/\langle p_T\rangle$ along with the linear fit, leads to a very large $\mu$ value where the curves can cross ($\mu\sim 3~$GeV).  There is no reason however to assume a linear fit function, so in Fig.~4 we also try a second order polynomial. The assumption of a second order polynomial function for $f$ allows the curves from different system sizes to cross at a much smaller value of $\mu$. Based on this fit, we find that the data is consistent with a critical point at $\mu\sim510~$MeV corresponding to a $\sqrt{s_{NN}}$ of 5.75 GeV.

\begin{figure}[htb]
\centering\mbox{
\includegraphics[width=0.7\textwidth]{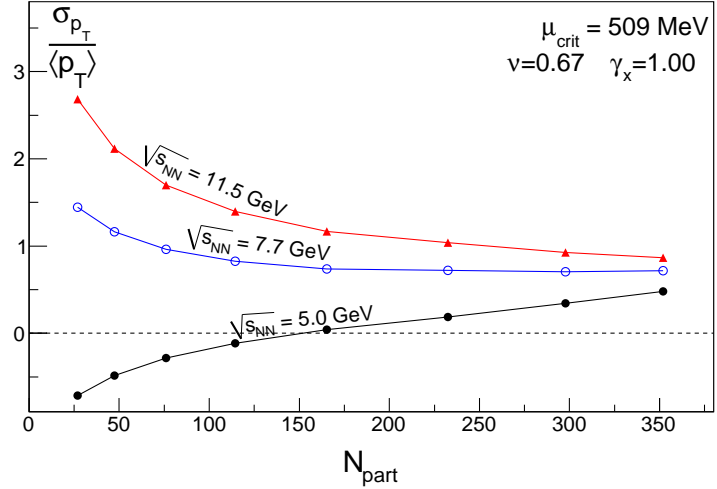}}
\caption[]{ The expected measurement of $\sigma_{p_T}/\langle p_T\rangle$ as a function 
of the number of participants at lower energies assuming the critical point is at $509$ MeV 
as extracted from the quadratic polynomial fit of STAR data.}
\label{predict}
\end{figure}

RHIC has also run at lower energies in order to search for a critical point in the Beam Energy 
Scan program.  Using the quadratic polynomial fit of STAR data (Fig. \ref{scaling-pol2}) and 
assuming the critical point is at $509$ MeV we can make 
predictions for $\sigma_{p_T}/\langle p_T\rangle$ at lower energies. 
We show this expectation as a function of the number of participants, $N_{\mathrm{part}}$, for three 
proposed beam energies: $11.5, 7.7$ and $5$ GeV in Fig.~5. Notice that the 
centrality dependence changes once one moves to the other side of the critical point.
This is a condition enforced by finite-size scaling which provides a generic signal for having 
reached the first-order phase transition side of the critical point.

Finite size effects are non negligible in heavy ion experiments and the mapping of the phase 
diagram for QCD they can produce. In fact, most thermodynamic quantities will be considerably 
shifted in the pseudocritical diagram that is actually probed, so that comparisons to results 
from the lattice in the thermodynamic limit should be made with caution. On the other hand, 
finite-size scaling techniques are simple and well defined in the case of heavy ion collisions. 
Even if it is hard to define the ideal scaling variable, it provides a pragmatic method to search 
for the critical point and investigate the universality properties of QCD. The fact that FSS 
prescinds from the knowledge of the details of the system under consideration, providing 
information about its criticality based solely on its most general features, makes it a very powerful 
tool for data analysis. From a very limited data set in energy spam, we have used FSS to exclude 
the presence of a critical point at small values of the baryonic chemical potential, below $450~$MeV. 
Besides, FSS is naturally fit to produce concrete predictions: once one assumes the existence 
of criticality, the scaling of correlation functions must be manifest in its vicinity. 

We are grateful to M. Chernodub, S.L.A. de Queiroz and \'A. M\'ocsy 
for fruitful discussions. This work was partially supported by CAPES-COFECUB (project 663/10), 
CNPq, FAPERJ and FUJB/UFRJ.



\begin{thebibliography}{99}

\bibitem{BES}
B.~I.~Abelev {\it et al.} [STAR Collaboration], 
STAR Note {\bf 0493} (2009); 
%
  M.~M.~Aggarwal {\it et al.}  [STAR Collaboration],
  arXiv:1007.2613 [nucl-ex].
  
\bibitem{critical}
  J.~Berges and K.~Rajagopal,
  Nucl.\ Phys.\  B {\bf 538}, 215 (1999); 
  %
  A.~M.~Halasz, A.~D.~Jackson, R.~E.~Shrock, M.~A.~Stephanov and J.~J.~M.~Verbaarschot,
  Phys.\ Rev.\  D {\bf 58}, 096007 (1998).
  
\bibitem{Cardy:1996xt} 
J.~L.~Cardy, 
\textit{Scaling and Renormalization in Statistical Physics} 
(Cambridge, 1996).

\bibitem{Stephanov:1998dy} 
M.~A.~Stephanov, K.~Rajagopal and E.~V.~Shuryak, 
Phys.\ Rev.\ Lett.\ \textbf{81}, 4816 (1998); 
Phys.\ Rev.\ D \textbf{60}, 114028 (1999). 

\bibitem{Stephanov:2008qz} 
M.~A.~Stephanov, 
Phys.\ Rev.\ Lett.\ \textbf{102}, 032301 (2009). 

\bibitem{Asakawa:2009aj}
  M.~Asakawa, S.~Ejiri and M.~Kitazawa,
  Phys.\ Rev.\ Lett.\  {\bf 103}, 262301 (2009).
%

\bibitem{Berdnikov:1999ph}
B.~Berdnikov and K.~Rajagopal, 
Phys.\ Rev.\ D \textbf{61}, 105017 (2000). 

\bibitem{Stephanov:2009ra}
  M.~A.~Stephanov,
  Phys.\ Rev.\  D {\bf 81}, 054012 (2010).
  
\bibitem{Gopie:1998qn} 
A.~Gopie and M.~C.~Ogilvie, 
Phys.\ Rev.\ D \textbf{59}, 034009 (1999). 

\bibitem{Bazavov:2007zz}
  A.~Bazavov and B.~A.~Berg,
  Phys.\ Rev.\  D {\bf 76}, 014502 (2007).

\bibitem{finite-NJL} 
O.~Kiriyama and A.~Hosaka, 
Phys.\ Rev.\ D \textbf{67}, 085010 (2003); 
%
O.~Kiriyama, T.~Kodama and T.~Koide, 
arXiv:hep-ph/0602086; 
%
L.~M.~Abreu, M.~Gomes and A.~J.~da Silva, 
Phys.\ Lett.\ B \textbf{642}, 551 (2006); 
%
  L.~M.~Abreu, A.~P.~C.~Malbouisson, J.~M.~C.~Malbouisson, A.~E.~Santana,
  Nucl.\ Phys.\  {\bf B819}, 127-138 (2009); 
%
  L.~M.~Abreu, A.~P.~C.~Malbouisson, J.~M.~C.~Malbouisson,
  Phys.\ Rev.\  {\bf D83}, 025001 (2011).

\bibitem{Braun:2004yk} 
J.~Braun, B.~Klein and H.~J.~Pirner, 
Phys.\ Rev.\ D \textbf{71}, 014032 (2005); 
Phys.\ Rev.\ D \textbf{72}, 034017 (2005); 
J.~Braun, B.~Klein, H.~J.~Pirner and A.~H.~Rezaeian, 
Phys.\ Rev.\ D \textbf{73}, 074010 (2006). 

\bibitem{Yamamoto:2009ey}
  N.~Yamamoto and T.~Kanazawa,
  Phys.\ Rev.\ Lett.\  {\bf 103}, 032001 (2009).

\bibitem{Palhares:2009tf}
  L.~F.~Palhares, E.~S.~Fraga and T.~Kodama,
  arXiv:0904.4830 [nucl-th] (J.\ Phys.\ G, in press);
%
  PoS {\bf CPOD2009}, 011 (2009);
  %
  J.\ Phys.\ G {\bf G37}, 094031 (2010).
  
\bibitem{Spieles:1997ab} 
C.~Spieles, H.~Stoecker and C.~Greiner, 
Phys.\ Rev.\ C \textbf{57}, 908 (1998). 

\bibitem{Fraga:2003mu} 
E.~S.~Fraga and R.~Venugopalan, 
Physica A \textbf{345}, 121 (2004). 

\bibitem{GellMann:1960np} M.~Gell-Mann and M.~Levy, 
Nuovo Cim.\ \textbf{16}, 705 (1960). 

\bibitem{Adams:2005dq}
 J.~Adams {\it et al.}  [STAR Collaboration],
 Nucl.\ Phys.\  A {\bf 757}, 102 (2005).

\bibitem{fisher} 
M.~E.~Fisher, 
in \textit{Critical Phenomena, Proc. 51st Enrico Fermi Summer School, Varena}, 
ed. M. S. Green (Academic Press, NY, 1972); 
M.~E.~Fisher and M.~N.~Barber, 
Phys.\ Rev.\ Lett.\ \textbf{28}, 1516 (1972). 

\bibitem{Brezin:1981gm} E.~Brezin, 
J. Physique \textbf{43}, 15 (1982). 

\bibitem{Brezin:1985xx} E.~Brezin and J.~Zinn-Justin, 
Nucl.\ Phys.\ B \textbf{257}, 867 (1985). 

\bibitem{Fraga:2011hi}
  E.~S.~Fraga, L.~F.~Palhares and P.~Sorensen,
  arXiv:1104.3755 [hep-ph] (Phys. Rev. C, in press).
  
\bibitem{amit} 
D. Amit, 
\textit{Field Theory; The Renormalization Group and Critical Phenomena} 
(World Scientific, 2005).

\bibitem{Adams:2005ka}
  J.~Adams {\it et al.}  [STAR Collaboration],
  Phys.\ Rev.\  C {\bf 72}, 044902 (2005).

\end{thebibliography}
\end{document}